\documentclass[10pt,conference]{IEEEtran}

\usepackage{cite}
\usepackage{url}
\usepackage{amsmath,amssymb,amsfonts}
\usepackage{algorithmic}
\usepackage{graphicx}
\usepackage{textcomp}
\usepackage{xcolor}
\usepackage{xspace}
\usepackage{balance}

\usepackage{makecell}
\usepackage{listings}
\usepackage[utf8]{inputenc}
\usepackage{colortbl}
\usepackage{multirow}
\usepackage{booktabs}

\usepackage[dvipsnames]{xcolor}
\usepackage{tikz}
\usetikzlibrary{positioning, backgrounds, calc}

\definecolor{bgWhite}{HTML}{FFFFFF}
\definecolor{bgSummary}{HTML}{D5D9DC}
\definecolor{bdSummary}{HTML}{7B848C}
\definecolor{bgRef}{HTML}{BCE4EA}
\definecolor{bdRef}{HTML}{369DB0}
\definecolor{bgRepro}{HTML}{BDCFF3}
\definecolor{bdRepro}{HTML}{3B71E1}
\definecolor{bgObs}{HTML}{ECC1D4}
\definecolor{bdObs}{HTML}{C83D7D}
\definecolor{bgExp}{HTML}{C6DFCC}
\definecolor{bdExp}{HTML}{287A41}

\newcommand{\ie}{\emph{i.e.,}\xspace}
\newcommand{\eg}{\emph{e.g.,}\xspace}

\newcommand{\etal}{\emph{et~al.}\xspace}
\newcommand{\secref}[1]{Section~\ref{#1}\xspace}

\newcommand{\tabref}[1]{Table~\ref{#1}\xspace}

\newcommand{\swebench}{SWE-bench\xspace}
\newcommand{\swebenchverified}{SWE-bench Verified\xspace}
\definecolor{lightlightgrey}{RGB}{233, 236, 239}

\newboolean{showcomments}
\setboolean{showcomments}{true}
\ifthenelse{\boolean{showcomments}}
  {\newcommand{\nb}[2]{
    \fbox{\bfseries\sffamily\scriptsize#1}
    {\sf\small$\blacktriangleright$\textit{#2}$\blacktriangleleft$}
   }
  }
  {\newcommand{\nb}[2]{}
  }

\def\BibTeX{{\rm B\kern-.05em{\sc i\kern-.025em b}\kern-.08em
    T\kern-.1667em\lower.7ex\hbox{E}\kern-.125emX}}

\begin{document}

\title{Writing Bug Reports for Software Repair Agents: What Information Matters Most?}

\author{
\IEEEauthorblockN{
Vincenzo Luigi Bruno\IEEEauthorrefmark{1}\IEEEauthorrefmark{2}, 
Alessandro Giagnorio\IEEEauthorrefmark{1}\IEEEauthorrefmark{2}, 
Daniele Bifolco\IEEEauthorrefmark{1}\IEEEauthorrefmark{3}, \\
Leon Wienges\IEEEauthorrefmark{2}, 
Massimiliano Di Penta\IEEEauthorrefmark{3}, and 
Gabriele Bavota\IEEEauthorrefmark{2}
}

\IEEEauthorblockA{\IEEEauthorrefmark{2}\textit{USI Università della Svizzera italiana}, Lugano, Switzerland \\
Email: \{vincenzo.luigi.bruno, alessandro.giagnorio, leon.wienges, gabriele.bavota\}@usi.ch}

\IEEEauthorblockA{\IEEEauthorrefmark{3}\textit{University of Sannio}, Benevento, Italy \\
Email: d.bifolco@studenti.unisannio.it, dipenta@unisannio.it}

\IEEEauthorblockA{\IEEEauthorrefmark{1}These authors contributed equally to this research.}
}

\maketitle
\thispagestyle{plain}

\begin{abstract}
Software development is increasingly moving toward agentic-first workflows.
This includes AI agents responsible for generating initial fixes for submitted issue reports.
In this setting, issue reports are no longer merely documentation for human maintainers; they become the primary task specification for the agent. However, little is known about how such reports should be written to maximize the agent's chances of producing a correct fix. We study what makes a bug report agent-ready. Starting from the \swebenchverified benchmark (\ie a collection of 500 real repository issues with human-written gold patches and test suites for evaluating generated fixes) we manually classify each issue by change type (\eg bug fix \emph{vs} refactoring) and annotate each sentence with its information type, such as observed behavior, expected behavior, reproduction steps, localization cues, and suggested fixes. We focus on the 441 issues representing bug reports, and we run on them mini-swe-agent using three LLM backbones (\ie GPT-5-mini, MiniMax M2.5, and Gemini 3 Flash). We then fit a binomial regression model to estimate the incremental association between each information type and agent success, controlling for confounding factors. Our results suggest that agentic-first reports benefit most from information that narrows the agent's search and repair space. Suggested fixes, expressed either in code or natural language, show the strongest positive associations with pass probability, while localization cues show positive but non-significant associations. An ablation study removing selected information types confirms that agents benefit less from information traditionally useful to humans, such as reproduction steps, and more from sentences that expose a repair direction. 
\end{abstract}


% !TEX root = ../main.tex
%%%%%%%%%%%%%%%%%%%%%%%%%%%%%%%%%%%%%%%%
%%%%%%%%%%%%%%%%%%%%%%%%%%%%%%%%%%%%%%%%
\section{Introduction} \label{sec:intro}
%%%%%%%%%%%%%%%%%%%%%%%%%%%%%%%%%%%%%%%%
%%%%%%%%%%%%%%%%%%%%%%%%%%%%%%%%%%%%%%%%
Software development workflows are increasingly incorporating LLM-based agents that can inspect repositories, reason over issues, modify code, and propose patches \cite{Zhang:2024,Ruan:icse2025,Ma:fse2025}. In these workflows, the human developer is no longer always the first actor. For example, in the context of bug fixing, a common interaction pattern is an agentic-first approach: an issue is assigned to an AI agent, the agent attempts to generate a patch, and the human developer reviews the result before accepting, modifying, or rejecting it. This shift changes the role of issue reports. So far, bug reports have been written primarily for human maintainers. A useful report helps a developer understand what went wrong, reproduce the problem, locate the relevant code, and reason about a possible fix \cite{Zimmermann:tse2010}. In an agentic-first workflow, however, the issue report also becomes the task specification for an automated repair process. 

The agent receives the issue as its initial description of the problem and must decide where to look, what behavior to reproduce, and how to construct a patch. Therefore, the practical question is no longer only \emph{what makes a bug report useful for humans} \cite{Zimmermann:tse2010}, but also \emph{what makes a bug report effective for agents}.

The distinction between human-oriented and agent-oriented issue quality is relevant even when evaluating agentic solutions for automated bug fixing, which are typically assessed through repository-level benchmarks. The latter (\eg \swebenchverified \cite{jimenez2024swebench}) are collections of real GitHub issues from open-source repositories, with accompanying gold patches and tests to assess automatically generated patches. However, the issues themselves vary considerably in the type and amount of information they contain. 
Some reports describe only symptoms, while others include reproduction steps, environment details, references to related discussions, affected code, and possible fixes. When an agent succeeds, it is unclear whether this is due to strong repository-level reasoning, to highly informative issue content, or to a combination of both. Understanding this distinction is essential to designing better agentic-first issue templates, interpreting benchmark results accurately, and writing issues that agents can act upon effectively.

In this paper, we investigate what types of information in bug reports are associated with the success of LLM-based software repair agents. We start with the 500 issues in \swebenchverified, inspecting them to retain only the 441 that correspond to defects, excluding 59 tasks that represent feature implementations or refactorings. We then manually annotate each issue at the character level using a taxonomy of bug-report information types coming from the existing literature \cite{Zimmermann:tse2010,Soltani:emse2020,Li:tse2023,Li:jss2023,sulun:tosem2024}. The taxonomy includes bug summary, observed behavior, expected behavior, reproduction steps, environment details, references to external resources (\eg other issues, Stack Overflow discussions), localization information at different granularity (\ie lines/functions/files likely affected by the bug), suggested fixes in code or natural language, and current workarounds. This annotation allows us to characterize issue reports as structured inputs to agents rather than as undifferentiated natural-language prompts. 

To assess the relation between issue information and agent performance, we run mini-swe-agent \cite{minisweagent} on each of the 441 bug reports using three LLM backbones: GPT-5-mini \cite{gpt5mini}, MiniMax M2.5 \cite{minimax_m25}, and Gemini 3 Flash \cite{gemini_3_flash}. We then fit a binomial regression model \cite{McCullagh:glm1989} estimating the incremental association between each information type and the odds of success, while controlling for issue length (\ie number of words composing the issue), gold-patch complexity (\ie the number of code lines changed by the original developer to fix the bug), and code mentions (\ie does the issue mention at least a code snippet). We found that suggested fixes show the strongest positive association with successful patches, both when provided as code and when expressed in natural language, while localization cues show positive but non-significant associations. 

These results suggest that agent-readiness is not simply traditional bug-report quality under a new name. Reports that merely document the existence of a failure are not enough; the information most closely associated with agent success is that which operationalizes the repair task. Also, agents do not simply benefit from more context (\ie longer issue descriptions). Indeed, specific types of information (\eg observed behavior) are not helpful for the agents. Instead, they benefit from concise, task-directive context. 

To complement the observational model with a more controlled assessment of individual report components, we perform a sentence-level ablation study. We identify a subset of 65 information-complete bug reports, \ie reports containing at least one text span for each information type considered in our study. For each issue, we then construct multiple variants by selectively removing one information type at a time (\eg localization cues, suggested fixes), as well as selected combinations of information (\eg both localization cues and suggested fixes). The results show that removing information considered important for human bug reporting, such as reproduction steps and expected behavior \cite{Zimmermann:tse2010,Laukannen:esem2011,Soltani:emse2020}, does not significantly reduce agent effectiveness when more operational cues remain available. In contrast, removing suggested fixes significantly reduces success, and the strongest degradation appears when they are removed together with localization cues. This suggests that agents benefit most from information that exposes a plausible repair direction.

The implication for agentic-first development is therefore concrete: issue reports should be written less as exhaustive narratives and more as repair specifications that tell the agent what failed, where to start looking, and what kind of change may be needed.

In summary, this paper makes the following contributions:

\begin{enumerate}
	\item We introduce the notion of agentic-first issue reports: bug reports written not only for human understanding but also to support AI agents attempting a repair. 
	
	\item We provide character-level annotations of information types for \swebenchverified, including behavioral information, environment details, references, localization cues, suggested fixes, and workarounds. Such a dataset can be used for research going beyond the objective of this work, \eg training and evaluating automated approaches to extract and classify information from bug reports, assessing their information completeness, and developing techniques to automatically enrich reports with information useful for human developers and AI-based bug-fixing agents.
	
	\item We empirically analyze the association between issue information and agent success across three LLM backbones and repeated executions of mini-swe-agent.
	
	\item Our results provide initial empirical evidence for the shift towards agentic-first issue reports, suggesting that future issue templates should explicitly consider the information needs of software repair agents.
	
\end{enumerate}

Code and data used in our study are publicly available \cite{replication}.
% !TEX root = ../main.tex

%%%%%%%%%%%%%%%%%%%%%%%%%%%%%%%%%%%%%%%%
%%%%%%%%%%%%%%%%%%%%%%%%%%%%%%%%%%%%%%%%
\section{Study Design} \label{sec:design}
%%%%%%%%%%%%%%%%%%%%%%%%%%%%%%%%%%%%%%%%

%%%%%%%%%%%%%%%%%%%%%%%%%%%%%%%%%%%%%%%%

The \emph{goal} of this study is to analyze bug reports to determine which of their parts help agentic AI to fix bugs. The \emph{quality focus} relates to the content of issues that, while largely investigated with respect to human maintainers \cite{Zimmermann:tse2010,Soltani:emse2020,Li:tse2023,Li:jss2023,sulun:tosem2024}, has received little attention in terms of the information content useful for agentic AI \cite{meng2026llm,ehsani2026llms}. The \emph{perspective} is of researchers wanting to develop approaches to improve issue report quality and enhance multi-agent systems for bug fixing. The \emph{context} consists of 500 issues from \swebenchverified \cite{jimenez2024swebench}, an agentic AI architecture (mini-swe-agent \cite{minisweagent}) and three LLMs: MiniMax M2.5 \cite{minimax_m25}, GPT-5-mini \cite{gpt5mini}, and Gemini 3 Flash \cite{gemini_3_flash}.
We address the following research question:

\begin{quote}
	\textbf{RQ:} What type of information contained in bug reports better correlates with agentic AI's ability to automatically fix bugs?
\end{quote}

\subsection{Benchmark Identification and Annotation}
We describe how we create a dataset of bug reports with character-level annotations of the information they contain.

\subsubsection{Selected Benchmark}
We leverage \swebenchverified \cite{jimenez2024swebench} as the starting point for our study. \swebenchverified is a human-validated subset of the \swebench benchmark, and it is composed of 500 real issues mined from open-source repositories. Each issue is paired with developer-written patches (gold patch) and test suites for evaluating generated patches. The input to the agent is (i) the repository state preceding the human fix and (ii) the corresponding issue report. The expected output is a patch that passes the benchmark tests. 

\begin{table*}[t]
	\centering
	\scriptsize
	\caption{Types of information labeled in the inspected bug reports.\vspace{-0.3cm}}
	\label{tab:information}
	\begin{tabular}{llll}
		\toprule
		\textbf{Info Type} & \textbf{Levels} & \textbf{Description} & \textbf{References}\\\midrule
		Summary & - & A high-level description of what the issue is & \cite{Zimmermann:tse2010, Li:tse2023, Li:jss2023, sulun:tosem2024} \\
		Observed behavior & - & What actually happened, including the manifested problem or failure & \cite{Zimmermann:tse2010,Li:tse2023,sulun:tosem2024} \\
		Expected behavior & - & What the reporter expected to happen instead & \cite{Zimmermann:tse2010, Li:tse2023, Li:jss2023, sulun:tosem2024} \\
		Reference & - & An external pointer to either other issues, project's resources, or Websites &  \cite{Li:tse2023,Li:jss2023,sulun:tosem2024} \\
		Reproduction steps & - & Step-by-step instructions for reproducing the issue & \cite{Zimmermann:tse2010, Soltani:emse2020, Li:tse2023, Li:jss2023, sulun:tosem2024} \\
		Environment & - &  Context where the issue occurs (\eg hardware, operating system) & \cite{Li:tse2023,Li:jss2023,sulun:tosem2024}  \\
		Affected area & line $|$ function $|$ file $|$ generic & Where in the system the problem occurs & \cite{Zimmermann:tse2010,Li:tse2023} \\
		Suggested fix & code $|$ natural language & A proposed fix, either as a code snippet/diff or in natural language & \cite{Soltani:emse2020,Li:tse2023,Li:jss2023} \\
		Current workaround & - & A temporary workaround until a full fix exists & \cite{sulun:tosem2024} \\\bottomrule
	\end{tabular}
	\vspace{-0.4cm}
\end{table*}

\subsubsection{Defining the Information Types to Annotate}
To support the manual annotation of issue sentences, we reviewed the existing literature to derive a set of information types commonly found in issue descriptions. We identified five articles that, either directly or indirectly, characterize the information content of issue reports \cite{Zimmermann:tse2010,Soltani:emse2020,Li:tse2023,Li:jss2023,sulun:tosem2024}. We inspected these articles and identified nine information types, reported in \tabref{tab:information} together with a short description, the references documenting each type, and, when applicable, a set of possible levels. For instance, \emph{location} may identify the code affected by the bug with different degrees of precision, ranging from an exact line to a generic description of the affected code area. During this process, we harmonized terminology and granularity across papers. This is because different studies sometimes use different terms for the same information type, or organize similar information at different levels of detail. Also, when finer-grained distinctions were not needed for our study, we grouped related elements into broader categories. For example, Zimmermann \etal \cite{Zimmermann:tse2010} distinguish among \emph{hardware}, \emph{software/project version}, and \emph{operating system/platform}. We merge these elements into the broader category \emph{environment}, following the terminology adopted by Li \etal \cite{Li:jss2023} and S\"ul\"un \etal \cite{sulun:tosem2024}.  We excluded information types that, although documented in prior work, are not relevant to our study because they are unlikely to affect an agent's ability to automatically fix a bug. 

Specifically, we excluded \emph{issue type/label} \cite{Li:jss2023}, since our analysis focuses exclusively on bug reports, as explained later. We also excluded \emph{assignee} \cite{Li:jss2023}, \emph{severity} \cite{Zimmermann:tse2010,Li:jss2023}, and \emph{knowledge level of the reporter} \cite{Li:jss2023}, since unrelated to the information an agent needs to generate a patch. Our replication package includes a spreadsheet summarizing the information types documented in the five inspected works \cite{Zimmermann:tse2010,Soltani:emse2020,Li:tse2023,Li:jss2023,sulun:tosem2024} and mapping them to the information types we consider \cite{replication}. While most information types in \tabref{tab:information} are self-explanatory, the distinction between \emph{summary} and \emph{observed behavior} deserves clarification. We use \emph{summary} for sentences that provide a high-level statement of the defect, describing the problem in general terms. We use \emph{observed behavior} for content reporting the concrete manifestation of the bug, \eg wrong outputs.

\subsubsection{Annotation Process}
Each of the 500 issues was independently assigned to two authors, who acted as annotators. Each annotator performed two tasks: (i) classifying the type of issue (\eg bug report, feature request), and (ii) annotating its textual content using the labels reported in \tabref{tab:information}. 

The issue classification was mainly guided by the labels assigned by the original developers, when available, and by our interpretation of the issue content otherwise. We classified 441 issues as bug reports, 41 as feature requests, and 18 as refactoring requests. Only 11 conflicts arose in this classification, with 8 being \emph{feature request} \emph{vs} \emph{bug} and 3 \emph{refactoring request} \emph{vs} \emph{bug} cases. We resolved conflicts via open discussion.

For annotating textual content, we developed a web application available in our replication package \cite{replication}. The application displayed the issue text and allowed annotators to select text spans with the mouse and assign them a label from \tabref{tab:information}. Text selection was supported at the character level, enabling annotators to mark only the fragments that actually conveyed a given information type. For example, in the sentence ``\emph{after upgrading to Python 3.11, calling \texttt{parse()} raises a \texttt{ValueError}, which is quite annoying because it blocks our nightly pipeline}'', ``\emph{Python 3.11}'' could be annotated as \emph{environment}, and only ``\emph{calling \texttt{parse()} raises a \texttt{ValueError}}'' as \emph{observed behavior}. The remaining text expresses the reporter's frustration and the consequence for their workflow, but does not directly describe information needed to characterize or fix the bug; therefore, it would not be assigned any of our labels.

After two authors had annotated all issues, conflicts were resolved by a third author who had not participated in the initial labeling. We assessed the agreement between the two annotators across four possible scenarios. 

The first two are agreement cases: (i) neither annotator selected the character as part of information (\ie above referred to as noise); and (ii) both annotators selected the character and assigned the same label. The last two are instead disagreement cases: (iii) only one annotator selected the character for labeling; and (iv) both annotators selected the character but assigned different labels.
Overall, the annotators agreed on 78.6\% of the issues' characters.

We further assess annotator agreement by computing Cohen's $k$ \cite{kappa} for the presence or absence of each label at the issue level (\tabref{tab:label_distribution}), with a median $k$ of 0.57 across labels, indicating moderate agreement. Since $k$ is strongly affected by label prevalence, we also report raw agreement. For near-universal or rare labels, such as summary and file-level localization, raw agreement remains high despite a low $k$ (\eg 96.6\% raw agreement for ``\emph{summary}'', despite the $k$ close to zero).

Finally, a third author resolved conflicts by inspecting both annotations and, when needed, discussing the rationale behind the original labeling decisions with the annotators.  We release the resulting labeled dataset in our replication package \cite{replication}. It covers all 500 issues in \swebench Verified and includes 3,752 annotations, averaging 7.5 per issue. 

\subsection{Agent Architecture and Selected LLMs}
We use mini-swe-agent \cite{minisweagent} as the agentic architecture for our experiments. This agent has been used to automate several code-related tasks \cite{team2026yet, baek2026artisan} and has demonstrated strong performance on \swebenchverified \cite{xia2025live}. Unlike other agent frameworks, mini-swe-agent does not rely on external tools; instead, it interacts with the environment exclusively through bash commands. At each iteration, the model is first prompted to reason about the next step, then to generate one or more bash commands to implement that step. The commands are subsequently executed in an isolated environment, and their outputs are returned to the model. Thus, each model iteration \emph{i} can be represented as a tuple $(T_i, (A_{i,1}, \dots, A_{i,n}), (O_{i,1}, \dots, O_{i,n}))$, where $T_i$ is the model's reasoning of the next step, $(A_{i,1} \dots A_{i,n})$ is the sequence of commands to execute in the shell, and $(O_{i,1}, \dots, O_{i,n})$ are the resulting outputs.

As subject LLMs, we select three of the top-performing models from the \swebench leaderboard \cite{swebenchleaderboard}, one being an \emph{open-weights model} (\ie MiniMax M2.5 \cite{minimax_m25}) and two \emph{closed-source} LLMs (\ie GPT-5-mini \cite{gpt5mini}, and Gemini 3 Flash \cite{gemini_3_flash}).
For all experiments, we set the maximum number of agent steps to 75, following previous work \cite{yang2025swesmith}, and set the \emph{reasoning effort} to \emph{medium} to contain inference costs. 

On the 8,649 runs we performed (details in \secref{sub:procedure}), the agents, on average, performed 36.6 steps (median=32.0). Of the 3,068 unsuccessful runs, 2,075 submitted an incorrect patch, 820 reached the step limit, and the remaining 173 submitted an empty patch, produced a patch that could not be evaluated, or stopped because of API errors \cite{replication}. All these runs are equally treated as ``unsuccessful'', independently from the reasons behind them.

\subsection{Empirical Study Procedure \& Data Analysis} 
\label{sub:procedure}

We detail the two procedures performed to answer our RQ.

\subsubsection{Observational Analysis on All Bug Reports}
\label{subsec:importance}
We first collect four issue-level features used as control variables in our models: the number of words in the issue description (\emph{Issue length}), a boolean indicator capturing whether the issue contains code snippets (\emph{Code mention}), the size of the gold-patch diff measured as the number of added and deleted lines (\emph{Gold patch size}), and the task difficulty (\emph{Difficulty}). We encode difficulty as 1 for issues estimated to take less than 15 minutes to fix, 2 for issues requiring between 15 minutes and 1 hour, and 3 for issues requiring over 1 hour. This information is provided by \swebenchverified as a human-estimated difficulty label. When \swebenchverified was created, three software developers were asked to estimate how long an experienced software engineer would take, after a few hours of familiarization with the codebase, to understand the issue, identify a solution, and implement the fix. 

Using the 441 bug reports in the selected benchmark, we first inspect the relationships among the four issue-level control variables by relying on a redundancy analysis using the \emph{redun} function from the \emph{Hmisc} \cite{hmisc} \emph{R} \cite{R} package. We consider a variable redundant when more than 80\% of its variance can be explained by the remaining variables. This analysis indicates that none of the four variables is redundant; therefore, we retain all of them in the statistical models. We then complement these issue-level controls with a set of binary variables, one for each information label, indicating whether the corresponding information type appears in the issue report.

We then define the dependent variable as an indicator of whether an agent successfully resolves each issue. To obtain this outcome, we use each selected LLM, one at a time, as the backbone of mini-swe-agent and run it on the 441 bug-report issues from \swebenchverified. For each run, we record whether the generated patch passes the task-specific test suite. To account for the stochastic nature of LLM-based agents, we execute each LLM three times per issue, yielding 3,969 observations in total (3 models $\times$ 3 runs $\times$ 441 issues).

Using the variables described above, we fit a mixed-effect binomial model \cite{McCullagh:glm1989} to assess which characteristics of a bug report are associated with the agent's success. Before fitting the model, we log-transform and standardize both issue length and gold-patch size to mitigate the effect of their skewed distributions. The model predictors include: (i) the information-type indicators, which represent our predictors of interest (\ie whether an issue contains a specific information type); (ii) the four issue-level metrics, used as control variables to adjust for issue characteristics (\eg \emph{Issue length}); and (iii) the LLM used as the agent backbone in each run, to account for differences in performance across models. Finally, we include the issue id as a random effect, allowing the model to account for issue-specific variation, such as differences in issue complexity. Since issues are nested in repositories, this term also absorbs repository-specific variation, such as differences in domain or codebase structure. Indeed, we tried adding a repository random effect as well but its estimated variance was zero, leaving results unchanged.
We refer to this model as the \emph{pooled model}, since it is fitted on all runs performed by all LLMs. In addition, we fit three separate \emph{per-LLM models}, each using only the runs produced with a specific LLM backbone. These models allow us to assess whether the findings observed in the pooled analysis are robust across different LLMs.

\subsubsection{Ablation Study on Information-Complete Bug Reports}
Next, we investigate how agents behave when specific types of information are missing from the issue description. This is done via an ablation study in which we manipulate information-complete issues by selectively removing specific information types. This allows us to compare the same set of bug reports under different treatments.

Testing all possible label combinations would yield an impractically large number of variants. We therefore group semantically related labels into a smaller set of macro-categories ($M$). 
The \emph{Bug location} category includes all labels indicating the affected area of the bug, whether at the line, function, file, or generic level. Similarly, \emph{Possible fix} groups suggested fixes expressed either as code or in natural language, since both provide the agent with a repair hypothesis. We keep \emph{summary} and \emph{observed behavior} as separate categories because they are present in almost all issues and represent two basic, yet distinct, components of a bug report. Instead, we group \emph{reproduction steps} and \emph{expected behavior} into \emph{Behavioral context}, since together they describe how the failure can be exercised and what the correct behavior should be. This grouping allows us to reduce the experimental space while preserving the main functional roles that issue information can play for an agent: describing the problem, exposing the failure, providing behavioral context, localizing the defect, and suggesting a repair direction.

From the 441 bug reports, we retain only those containing all five macro-categories. This filtering ensures that every treatment described below can be applied to the same set of tasks, enabling direct comparisons across variants. This process resulted in 65 \emph{complete} issues. For these issues, we first remove all text spans that were not labeled by any annotator, to minimize the influence of content not associated with any of our categories and better isolate the contribution of the labeled information. These unlabeled spans mostly consist of template text, comments, or details that do not provide task-relevant information. We refer to this cleaned version of the original issue as the ``\emph{no noise}'' treatment. We consider seven treatments in addition to ``\emph{no noise}''. The first five remove each macro-category independently. For instance, \emph{no localization} removes all spans providing localization cues. 

The remaining two treatments remove combinations of macro-categories designed to capture conditions of specific interest for our study. The first, \emph{no localization, no suggestions}, removes both localization cues and suggested fixes. Since these are the two most actionable information types, this treatment allows us to study how agents behave when the issue provides little direct guidance on where to look or how to intervene. The second, \emph{no behavioral context, no localization, no suggestions}, simulates issues that retain only the two information types appearing in almost all bug reports, namely \emph{summary} and \emph{observed behavior}. 

We create eight versions of each issue: the \emph{no noise} version, and one version for each of the seven treatments. Each treatment version is obtained by automatically removing the spans associated with the corresponding macro-category or combination of macro-categories. Since automatic removal can introduce grammatical or semantic issues, such as broken sentences or incoherent transitions, two authors independently reviewed each mutated task to verify that it remained meaningful after removal. When needed, they edited the text to restore coherence without changing its semantic content. As a concrete example, one manipulated issue contained the line of code likely responsible for the bug, followed by the sentence ``\emph{this line produces RuntimeWarning: [...]}''. In this case, ``\emph{this line}'' referred to the likely buggy line, while ``RuntimeWarning: [...]'' described the observed behavior. In the \emph{no localization} treatment, removing the buggy line made the sentence incoherent, since ``\emph{this line}'' no longer had a clear referent. We therefore rewrote the sentence as ``RuntimeWarning: [...]'', preserving the observed behavior while avoiding a broken flow in the issue description. Overall, the authors inspected 520 mutated issues (65 issues $\times$ 8 treatments), achieving a Cohen's $k$ inter-rater agreement of 0.96 on the need for modifying manipulated sentences. 
Disagreements were resolved through open discussion involving a third author.

Finally, we run the three LLMs on the newly generated task variants, repeating each run three times and recording whether the agent successfully resolves the task. Overall, this results in 4,680 agent runs (3 LLMs $\times$ 8 treatments $\times$ 3 repetitions $\times$ 65 tasks). Using these data, we fit a mixed-effect binomial regression model including the same issue-level control variables and random effects described in \secref{subsec:importance}, together with a categorical variable representing the treatment applied to the issue in each run. 
As for the observational analysis, we report both a \emph{pooled model}, fitted across all three LLMs, and three \emph{per-LLM models}, each fitted only to runs produced with a specific LLM.

\subsection{Models' Diagnostics and Metrics}
After fitting each mixed-effect binomial model, we perform a set of diagnostic checks to assess the reliability of the estimates. First, we inspect the variance of the issue-level random intercept. For the eight models we built (\ie two \emph{pooled} and six \emph{per-LLM}), its standard deviation ranges between 2.42 and 5.48, indicating that, even after controlling for all other variables, some issues remain easier/harder for the agent. 

Second, we assess whether the models suffer from a singular fit, occurring when the random-effect variance is close to zero. Singular fit was not detected for any of the built models.
Third, we assess overdispersion, which occurs when the model underestimates the data variability. We use the dispersion ratio and its associated test. A slight overdispersion (dispersion ratio below 1.05) was detected only in the GPT-5-mini and MiniMax M2.5 observational models. However, the magnitude of overdispersion is negligible.

Finally, we analyze the collinearity among predictors using variance inflation factors. Low values indicate that the model predictors provide sufficiently distinct information, while high values suggest that some coefficients may be unstable because the predictors overlap strongly. We use common interpretation thresholds: values below 3 indicate low or acceptable collinearity, which was the case for all variables in all models.

For each model, we report the effect of each predictor using the odds ratio (OR), which, for binomial models, equals $e^{estimate}$ and captures both the magnitude and the direction of its association with the dependent variable. An OR greater than 1 indicates a positive association with the probability of success, whereas an OR lower than 1 indicates a negative association. We complement ORs with $p$-values to assess statistical significance, using a threshold of 0.05, with Wald 95\% confidence intervals (CIs) \cite{Wald:tams1943}, which indicate the range in which the true OR is expected to lie with 95\% confidence.
% !TEX root = ../main.tex

\begin{table}[t]
	\centering
	\footnotesize
	\caption{\label{tab:label_distribution}Distribution of issue-content labels and inter-annotator agreement (Cohen's $k$ and raw agreement).\vspace{-0.4cm}}
	\setlength{\tabcolsep}{3pt}
	\begin{tabular}[t]{lrrrr}
		\toprule
		\textbf{Label} & \textbf{\# Issues} & \textbf{\% Issues} & \textbf{$k$} & \textbf{\% Agr.}\\
		\midrule
		\cellcolor{gray!10}{Summary} & \cellcolor{gray!10}{441} & \cellcolor{gray!10}{100.00} & \cellcolor{gray!10}{$-$0.02} & \cellcolor{gray!10}{96.6}\\
		Observed behavior & 432 & 97.96 & 0.12 & 92.5\\
		\cellcolor{gray!10}{Reproduction steps} & \cellcolor{gray!10}{375} & \cellcolor{gray!10}{85.03} & \cellcolor{gray!10}{0.59} & \cellcolor{gray!10}{88.0}\\
		Expected behavior & 290 & 65.76 & 0.70 & 85.7\\
		\cellcolor{gray!10}{Environment} & \cellcolor{gray!10}{212} & \cellcolor{gray!10}{48.07} & \cellcolor{gray!10}{0.88} & \cellcolor{gray!10}{94.1}\\
		Reference & 205 & 46.49 & 0.65 & 82.5\\
		\cellcolor{gray!10}{Suggested fix: natural language} & \cellcolor{gray!10}{84} & \cellcolor{gray!10}{19.05} & \cellcolor{gray!10}{0.55} & \cellcolor{gray!10}{87.8}\\
		Affected area: function & 79 & 17.91 & 0.34 & 81.4\\
		\cellcolor{gray!10}{Affected area: generic} & \cellcolor{gray!10}{78} & \cellcolor{gray!10}{17.69} & \cellcolor{gray!10}{0.27} & \cellcolor{gray!10}{79.1}\\
		Affected area: line & 62 & 14.06 & 0.70 & 93.2\\
		\cellcolor{gray!10}{Suggested fix: code} & \cellcolor{gray!10}{38} & \cellcolor{gray!10}{8.62} & \cellcolor{gray!10}{0.76} & \cellcolor{gray!10}{96.1}\\
		Current workaround & 31 & 7.03 & 0.57 & 94.8\\
		\cellcolor{gray!10}{Affected area: file} & \cellcolor{gray!10}{22} & \cellcolor{gray!10}{4.99} & \cellcolor{gray!10}{0.19} & \cellcolor{gray!10}{93.4}\\
		\bottomrule
	\end{tabular}
	\vspace{-0.4cm}
\end{table}

\section{Study Results} \label{sec:results}

\tabref{tab:label_distribution} reports the distribution of information types across the 441 annotated bug reports. The table shows that some information types appear in nearly all reports, whereas others are rare. As expected, \emph{summary} appears in all 441 bug reports and is therefore excluded from the statistical analyses, since its association with agent success cannot be estimated in the absence of variation. Similarly, \emph{observed behavior} appears in 97.96\% of the issues, leaving very few cases in which this information is absent. At the opposite end, some potentially useful information types occur only rarely, such as \emph{suggested fix} in the form of \emph{code} (8.62\%), \emph{current workaround} (7.03\%), and \emph{affected area} at \emph{file} granularity (4.99\%). 

Typically, such pieces of information are unlikely to appear in bug report descriptions; rather, they appear in follow-up comments, which are excluded from our experiment as unavailable at issue opening time. 
Note that for these highly imbalanced information types, the model has limited evidence to estimate their association with repair success, making it harder to detect statistically significant effects.
This is important to keep in mind while reading the results of the mixed-effect binomial models reported below.

We present the results of our two complementary analyses. The first (\secref{sub:observational}) is the observational analysis on all 441 bug reports to identify which information types are associated with agent success in naturally occurring issue reports. The second (\secref{sub:ablation}) is the controlled ablation study on the subset of 65 information-complete bug reports. 

\vspace{-0.3cm}

\begin{table}[!h]
	\centering
	\footnotesize
	\caption{\label{tab:agent_ready_all_llms}Mixed-effect binomial model predicting issue resolution success across all LLMs. Bold indicates p $<$ 0.05.\vspace{-0.55cm}}
	\centering
	\setlength{\tabcolsep}{4pt}
	\begin{tabular}[t]{lrrr}
		\toprule
		\textbf{Predictor} & \textbf{Odds ratio} & \textbf{95\% CI} & \textbf{p-value}\\
		\midrule
		\cellcolor{gray!10}{\textbf{Difficulty}} & \cellcolor{gray!10}{\textbf{0.50}} & \cellcolor{gray!10}{\textbf{[0.35, 0.72]}} & \cellcolor{gray!10}{\textbf{$<$0.001}}\\
		\textbf{Issue length (log words)} & \textbf{0.64} & \textbf{[0.43, 0.94]} & \textbf{0.023}\\
		\cellcolor{gray!10}{Code mention} & \cellcolor{gray!10}{0.97} & \cellcolor{gray!10}{[0.33, 2.89]} & \cellcolor{gray!10}{0.957}\\
		\textbf{Gold patch size (log diff)} & \textbf{0.37} & \textbf{[0.26, 0.52]} & \textbf{$<$0.001}\\
		\cellcolor{gray!10}{Observed behavior} & \cellcolor{gray!10}{0.47} & \cellcolor{gray!10}{[0.06, 3.86]} & \cellcolor{gray!10}{0.480}\\
		Expected behavior & 1.18 & [0.60, 2.30] & 0.635\\
		\cellcolor{gray!10}{Reproduction steps} & \cellcolor{gray!10}{1.89} & \cellcolor{gray!10}{[0.70, 5.12]} & \cellcolor{gray!10}{0.210}\\
		Environment & 0.82 & [0.42, 1.60] & 0.562\\
		\cellcolor{gray!10}{Affected area: line} & \cellcolor{gray!10}{2.12} & \cellcolor{gray!10}{[0.83, 5.38]} & \cellcolor{gray!10}{0.114}\\
		Affected area: function & 2.07 & [0.95, 4.54] & 0.069\\
		\cellcolor{gray!10}{Affected area: file} & \cellcolor{gray!10}{2.75} & \cellcolor{gray!10}{[0.66, 11.52]} & \cellcolor{gray!10}{0.167}\\
		Affected area: generic & 1.89 & [0.86, 4.15] & 0.113\\
		\cellcolor{gray!10}{\textbf{Suggested fix: code}} & \cellcolor{gray!10}{\textbf{3.97}} & \cellcolor{gray!10}{\textbf{[1.21, 13.01]}} & \cellcolor{gray!10}{\textbf{0.023}}\\
		\textbf{Suggested fix: natural language} & \textbf{3.16} & \textbf{[1.41, 7.07]} & \textbf{0.005}\\
		\cellcolor{gray!10}{Current workaround} & \cellcolor{gray!10}{0.78} & \cellcolor{gray!10}{[0.24, 2.53]} & \cellcolor{gray!10}{0.681}\\
		\textbf{Reference} & \textbf{0.49} & \textbf{[0.27, 0.89]} & \textbf{0.020}\\
		\cellcolor{gray!10}{\textbf{Model: GPT-5-mini}} & \cellcolor{gray!10}{\textbf{0.66}} & \cellcolor{gray!10}{\textbf{[0.52, 0.83]}} & \cellcolor{gray!10}{\textbf{$<$0.001}}\\
		\textbf{Model: Minimax M2.5} & \textbf{1.55} & \textbf{[1.23, 1.97]} & \textbf{$<$0.001}\\
		\bottomrule
	\end{tabular}
	\vspace{-0.4cm}
\end{table}

\begin{table*}[!t]
	\centering
	\footnotesize
	\caption{\label{tab:agent_ready_per_llm}Mixed-effect binomial models predicting issue resolution success for each LLM in isolation. Bold indicates p $<$ 0.05. Predictor names are in bold face when significant for all three LLMs.\vspace{-0.5cm}}
	\setlength{\tabcolsep}{3pt}
	\begin{tabular}[t]{lccc c ccc c ccc}
		\toprule
		\multicolumn{1}{c}{} 
		& \multicolumn{3}{c}{\textbf{GPT-5-mini}} 
		& \multicolumn{1}{c}{} 
		& \multicolumn{3}{c}{\textbf{Gemini 3 Flash}} 
		& \multicolumn{1}{c}{} 
		& \multicolumn{3}{c}{\textbf{MiniMax M2.5}} \\
		\cmidrule(lr){2-4} \cmidrule(lr){6-8} \cmidrule(lr){10-12}
		\textbf{Predictor} 
		& \textbf{OR} & \textbf{95\% CI} & \textbf{p-value} 
		& 
		& \textbf{OR} & \textbf{95\% CI} & \textbf{p-value} 
		& 
		& \textbf{OR} & \textbf{95\% CI} & \textbf{p-value}\\
		\midrule
		\cellcolor{gray!10}{\textbf{Difficulty}} & \cellcolor{gray!10}{\textbf{0.33}} & \cellcolor{gray!10}{\textbf{[0.19, 0.56]}} & \cellcolor{gray!10}{\textbf{$<$0.001}} & & \cellcolor{gray!10}{\textbf{0.39}} & \cellcolor{gray!10}{\textbf{[0.22, 0.70]}} & \cellcolor{gray!10}{\textbf{0.002}} & & \cellcolor{gray!10}{\textbf{0.46}} & \cellcolor{gray!10}{\textbf{[0.21, 0.99]}} & \cellcolor{gray!10}{\textbf{0.047}}\\
		Issue length (log words) & \textbf{0.49} & \textbf{[0.28, 0.87]} & \textbf{0.014} & & 0.60 & [0.33, 1.09] & 0.092 & & \textbf{0.42} & \textbf{[0.18, 0.96]} & \textbf{0.039}\\
		\cellcolor{gray!10}{Code mention} & \cellcolor{gray!10}{1.66} & \cellcolor{gray!10}{[0.35, 7.90]} & \cellcolor{gray!10}{0.527} & & \cellcolor{gray!10}{0.89} & \cellcolor{gray!10}{[0.18, 4.56]} & \cellcolor{gray!10}{0.894} & & \cellcolor{gray!10}{0.46} & \cellcolor{gray!10}{[0.05, 4.75]} & \cellcolor{gray!10}{0.517}\\
		\textbf{Gold patch size (log diff)} & \textbf{0.33} & \textbf{[0.20, 0.57]} & \textbf{$<$0.001} & & \textbf{0.33} & \textbf{[0.19, 0.59]} & \textbf{$<$0.001} & & \textbf{0.07} & \textbf{[0.03, 0.18]} & \textbf{$<$0.001}\\
		\cellcolor{gray!10}{Observed behavior} & \cellcolor{gray!10}{0.10} & \cellcolor{gray!10}{[0.01, 1.92]} & \cellcolor{gray!10}{0.127} & & \cellcolor{gray!10}{0.50} & \cellcolor{gray!10}{[0.02, 12.62]} & \cellcolor{gray!10}{0.674} & & \cellcolor{gray!10}{0.59} & \cellcolor{gray!10}{[0.01, 42.70]} & \cellcolor{gray!10}{0.807}\\
		Expected behavior & 1.51 & [0.58, 3.90] & 0.395 & & 1.02 & [0.37, 2.83] & 0.972 & & 1.25 & [0.31, 5.09] & 0.753\\
		\cellcolor{gray!10}{Reproduction steps} & \cellcolor{gray!10}{1.73} & \cellcolor{gray!10}{[0.41, 7.21]} & \cellcolor{gray!10}{0.454} & & \cellcolor{gray!10}{2.46} & \cellcolor{gray!10}{[0.55, 10.92]} & \cellcolor{gray!10}{0.238} & & \cellcolor{gray!10}{7.02} & \cellcolor{gray!10}{[0.83, 59.16]} & \cellcolor{gray!10}{0.073}\\
		Environment & 0.85 & [0.33, 2.19] & 0.736 & & 1.17 & [0.42, 3.25] & 0.768 & & 0.64 & [0.16, 2.59] & 0.532\\
		\cellcolor{gray!10}{Affected area: line} & \cellcolor{gray!10}{2.94} & \cellcolor{gray!10}{[0.80, 10.81]} & \cellcolor{gray!10}{0.104} & & \cellcolor{gray!10}{2.41} & \cellcolor{gray!10}{[0.61, 9.61]} & \cellcolor{gray!10}{0.211} & & \cellcolor{gray!10}{2.81} & \cellcolor{gray!10}{[0.44, 17.90]} & \cellcolor{gray!10}{0.273}\\
		Affected area: function & \textbf{3.77} & \textbf{[1.22, 11.69]} & \textbf{0.021} & & 1.97 & [0.60, 6.46] & 0.266 & & 3.51 & [0.71, 17.40] & 0.124\\
		\cellcolor{gray!10}{Affected area: file} & \cellcolor{gray!10}{5.58} & \cellcolor{gray!10}{[0.77, 40.27]} & \cellcolor{gray!10}{0.088} & & \cellcolor{gray!10}{3.02} & \cellcolor{gray!10}{[0.39, 23.47]} & \cellcolor{gray!10}{0.291} & & \cellcolor{gray!10}{1.41} & \cellcolor{gray!10}{[0.10, 20.29]} & \cellcolor{gray!10}{0.800}\\
		Affected area: generic & \textbf{3.13} & \textbf{[1.01, 9.65]} & \textbf{0.048} & & 3.23 & [0.96, 10.88] & 0.059 & & 2.24 & [0.44, 11.49] & 0.334\\
		\cellcolor{gray!10}{Suggested fix: code} & \cellcolor{gray!10}{\textbf{5.58}} & \cellcolor{gray!10}{\textbf{[1.10, 28.26]}} & \cellcolor{gray!10}{\textbf{0.038}} & & \cellcolor{gray!10}{\textbf{7.30}} & \cellcolor{gray!10}{\textbf{[1.27, 41.95]}} & \cellcolor{gray!10}{\textbf{0.026}} & & \cellcolor{gray!10}{3.71} & \cellcolor{gray!10}{[0.38, 36.67]} & \cellcolor{gray!10}{0.261}\\
		\textbf{Suggested fix: natural language} & \textbf{4.66} & \textbf{[1.45, 14.94]} & \textbf{0.010} & & \textbf{5.73} & \textbf{[1.65, 19.96]} & \textbf{0.006} & & \textbf{9.44} & \textbf{[1.77, 50.25]} & \textbf{0.009}\\
		\cellcolor{gray!10}{Current workaround} & \cellcolor{gray!10}{0.53} & \cellcolor{gray!10}{[0.10, 2.87]} & \cellcolor{gray!10}{0.464} & & \cellcolor{gray!10}{0.68} & \cellcolor{gray!10}{[0.11, 4.13]} & \cellcolor{gray!10}{0.677} & & \cellcolor{gray!10}{0.32} & \cellcolor{gray!10}{[0.02, 4.36]} & \cellcolor{gray!10}{0.394}\\
		Reference & \textbf{0.36} & \textbf{[0.15, 0.85]} & \textbf{0.019} & & \textbf{0.30} & \textbf{[0.12, 0.78]} & \textbf{0.013} & & 0.40 & [0.11, 1.40] & 0.150\\
		\bottomrule
	\end{tabular}
	\vspace{-0.3cm}
\end{table*}

\subsection{Observational Analysis on All Bug Reports}
\label{sub:observational}

\tabref{tab:agent_ready_all_llms} reports the pooled mixed-effect binomial model, in which the outcomes of all runs across the three LLM backbones are analyzed together. \tabref{tab:agent_ready_all_llms} shows, for each predictor, the estimated odds ratio (OR), 95\% confidence interval, and p-value. \tabref{tab:agent_ready_per_llm} reports the same information for models fitted separately for each LLM backbone.

The results in \tabref{tab:agent_ready_all_llms} provide a consistent message: agent-ready bug reports are not simply longer or more detailed reports, but reports that make the repair task more operational. In the pooled model, the strongest positive associations with successful repair are observed for information suggesting how the defect might be fixed. Both code-level (OR=3.97) and natural-language (OR=3.16) suggested fixes are associated with substantially higher odds of success, and they are the only information types with a significant positive association. This suggests that agents can leverage explicit repair intent, even when expressed as text without a suggested patch. 

Localization information points in the same direction, with ORs between 1.89 and 2.75 for all granularities (\ie line, function, file and generic affected area), but none of these associations is statistically significant. 

The generic \emph{code mention} variable is also not significant (OR=0.97). Behavioral information shows a similar pattern. Expected behavior (OR=1.18) and reproduction steps (OR=1.89) have ORs above one but are not statistically significant, and neither is observed behavior. This should not be interpreted as evidence that observed behavior is unimportant. Indeed, almost all bug reports contain observed behavior (\tabref{tab:label_distribution}), leaving little variation for the model to estimate its incremental contribution. Our ablation study in \secref{sub:ablation} will shed more light on the contribution of such labels, since we will explicitly manipulate them.

The negative associations for issue length, gold-patch size, and difficulty provide further insights. First, longer issues are associated with lower odds of success, suggesting that length may capture verbosity, ambiguity, or the intrinsic complexity of the reported defect, beyond the controls included in the model. Second, the fact that larger gold patches and greater difficulty are associated with lower success odds suggests that agent-ready reporting (\ie focusing on information useful for the agent) can make a task easier to approach, but it cannot fully compensate for defects requiring large/difficult changes. 

Finally, references are negatively associated with success. This result should be interpreted with caution. References are unlikely to be harmful per se; rather, they may serve as proxies for issues whose relevant context is distributed across external resources. Such issues may be harder for agents because the necessary repair rationale is not fully contained in the report itself. This suggests that from an agentic-first perspective, merely linking to additional context may be insufficient. When references are included, the report should also summarize the repair-relevant information from those references, so the agent does not need to infer which external details matter. That is, while a human can ``understand'' the purpose of a reference and selectively process it, the agent may require suitable instructions and/or contextual information.

The per-LLM models in \tabref{tab:agent_ready_per_llm} show that several of the findings discussed above are consistent across LLMs, while others depend on the specific backbone. 
The most robust result concerns natural-language suggested fixes, which are statistically significant across all three LLMs with odds ratios between 4.66 and 9.44. Code-level suggested fixes are significant for GPT-5-mini and Gemini 3 Flash, but not for MiniMax M2.5 ($p$-value = 0.261). This reinforces the importance of suggested fixes, independently of whether repair guidance is expressed in code or natural language.

Localization information has ORs above one for all LLMs and granularities, but it is  significant only for GPT-5-mini, at the function (OR=3.77) and generic (OR=3.13) level.

Other differences concern references, negatively associated with success for GPT-5-mini (OR=0.36) and Gemini 3 Flash (OR=0.30) but not for MiniMax M2.5.
These differences suggest that agent backbones may leverage issue information in different ways. Thus, project owners adopting agentic-first workflows should treat issue templates as agent-specific interfaces: they should evaluate which information their chosen agent needs and adapt reporting guidelines accordingly. 

Gold-patch size, issue length, and difficulty are negative across models, confirming the findings of the pooled model. Taken together, the observational analysis suggests that agent-ready reports are those that transform a bug report from a descriptive artifact into an actionable repair specification, mainly by exposing possible repair directions while avoiding verbosity that may reflect noise or task complexity. At the same time, because these associations are estimated on naturally occurring reports, they do not establish whether each information type is independently necessary for success. The controlled ablation study in \secref{sub:ablation} assesses how agent performance changes when specific components are missing.

\begin{table}[!t]
	\centering
	\scriptsize
	\caption{\label{tab:agent_ready_manipulation_all_llms}Ablation study: Mixed-effect binomial model predicting issue resolution success across all LLMs. Bold indicates p$<$0.05.\vspace{-0.5cm}}
	\setlength{\tabcolsep}{3pt}
	\begin{tabular}[t]{lrrr}
		\toprule
		\textbf{Predictor} & \textbf{OR} & \textbf{95\% CI} & \textbf{p-value}\\
		\midrule
		\cellcolor{gray!10}{Difficulty} & \cellcolor{gray!10}{0.63} & \cellcolor{gray!10}{[0.26, 1.53]} & \cellcolor{gray!10}{0.309}\\
		Issue length (log words) & 1.41 & [0.65, 3.08] & 0.387\\
		\cellcolor{gray!10}{Code mention} & \cellcolor{gray!10}{0.41} & \cellcolor{gray!10}{[0.03, 6.62]} & \cellcolor{gray!10}{0.532}\\
		\textbf{Gold patch size (log diff)} & \textbf{0.33} & \textbf{[0.14, 0.76]} & \textbf{0.009}\\
		\cellcolor{gray!10}{No behavioral context} & \cellcolor{gray!10}{0.82} & \cellcolor{gray!10}{[0.56, 1.21]} & \cellcolor{gray!10}{0.325}\\
		No localization & 0.78 & [0.53, 1.14] & 0.203\\
		\cellcolor{gray!10}{No observed behavior} & \cellcolor{gray!10}{0.87} & \cellcolor{gray!10}{[0.59, 1.29]} & \cellcolor{gray!10}{0.489}\\
		\textbf{No suggestions} & \textbf{0.65} & \textbf{[0.45, 0.95]} & \textbf{0.027}\\
		\cellcolor{gray!10}{No summary} & \cellcolor{gray!10}{1.15} & \cellcolor{gray!10}{[0.78, 1.72]} & \cellcolor{gray!10}{0.478}\\
		\textbf{No localization, no suggestions} & \textbf{0.42} & \textbf{[0.29, 0.61]} & \textbf{$<$0.001}\\
		\cellcolor{gray!10}{\textbf{No behavioral context, no localization, no suggestions}} & \cellcolor{gray!10}{\textbf{0.38}} & \cellcolor{gray!10}{\textbf{[0.27, 0.56]}} & \cellcolor{gray!10}{\textbf{$<$0.001}}\\
		\textbf{Model: GPT-5-mini} & \textbf{1.31} & \textbf{[1.05, 1.65]} & \textbf{0.017}\\
		\cellcolor{gray!10}{\textbf{Model: MiniMax M2.5}} & \cellcolor{gray!10}{\textbf{1.47}} & \cellcolor{gray!10}{\textbf{[1.17, 1.84]}} & \cellcolor{gray!10}{\textbf{$<$0.001}}\\
		\bottomrule
	\end{tabular}
	\vspace{-0.4cm}
\end{table}

\subsection{Ablation Study on Information-Complete Bug Reports}
\label{sub:ablation}
The ablation study compares variants of the same 65 information-complete issues after selectively removing specific information types. Thus, the underlying defect remains fixed, while the amount and type of information available to the agent changes. \tabref{tab:agent_ready_manipulation_all_llms} reports the results of the pooled model across all LLMs. In this analysis, each treatment denotes a specific manipulation of the original issue text. For example, the treatment ``\emph{no localization}'' indicates that all localization cues were removed from the report, regardless of their granularity. 

\begin{table*}[!t]
	\centering
	\footnotesize
	\caption{\label{tab:ablation_by_model}Mixed-effect binomial models predicting issue resolution success for each LLM in isolation under report-content manipulation. Estimates are odds ratios. Bold indicates p $<$ 0.05. Predictor names are in bold face when significant for all three LLMs.\vspace{-0.5cm}}
	\setlength{\tabcolsep}{3pt}
	\begin{tabular}[t]{lccc c ccc c ccc}
		\toprule
		\multicolumn{1}{c}{}
		& \multicolumn{3}{c}{\textbf{GPT-5-mini}}
		& \multicolumn{1}{c}{}
		& \multicolumn{3}{c}{\textbf{Gemini 3 Flash}}
		& \multicolumn{1}{c}{}
		& \multicolumn{3}{c}{\textbf{MiniMax M2.5}} \\
		\cmidrule(lr){2-4} \cmidrule(lr){6-8} \cmidrule(lr){10-12}
		\textbf{Predictor}
		& \textbf{OR} & \textbf{95\% CI} & \textbf{p-value}
		&
		& \textbf{OR} & \textbf{95\% CI} & \textbf{p-value}
		&
		& \textbf{OR} & \textbf{95\% CI} & \textbf{p-value}\\
		\midrule
		\cellcolor{gray!10}{Difficulty}
		& \cellcolor{gray!10}{0.60}
		& \cellcolor{gray!10}{[0.26, 1.40]}
		& \cellcolor{gray!10}{0.236}
		&
		& \cellcolor{gray!10}{0.65}
		& \cellcolor{gray!10}{[0.20, 2.13]}
		& \cellcolor{gray!10}{0.480}
		&
		& \cellcolor{gray!10}{0.65}
		& \cellcolor{gray!10}{[0.19, 2.20]}
		& \cellcolor{gray!10}{0.485}\\
		
		Issue length (log words)
		& 1.52
		& [0.72, 3.20]
		& 0.270
		&
		& 1.41
		& [0.49, 4.07]
		& 0.524
		&
		& 1.35
		& [0.46, 3.92]
		& 0.584\\
		
		\cellcolor{gray!10}{Code mention}
		& \cellcolor{gray!10}{0.26}
		& \cellcolor{gray!10}{[0.02, 3.68]}
		& \cellcolor{gray!10}{0.322}
		&
		& \cellcolor{gray!10}{1.52}
		& \cellcolor{gray!10}{[0.04, 57.53]}
		& \cellcolor{gray!10}{0.821}
		&
		& \cellcolor{gray!10}{0.21}
		& \cellcolor{gray!10}{[0.00, 10.90]}
		& \cellcolor{gray!10}{0.443}\\
		
		Gold patch size (log diff)
		& \textbf{0.34}
		& \textbf{[0.15, 0.77]}
		& \textbf{0.009}
		&
		& 0.37
		& [0.12, 1.12]
		& 0.078
		&
		& \textbf{0.29}
		& \textbf{[0.09, 0.93]}
		& \textbf{0.037}\\
		
		\cellcolor{gray!10}{No behavioral context}
		& \cellcolor{gray!10}{0.85}
		& \cellcolor{gray!10}{[0.44, 1.63]}
		& \cellcolor{gray!10}{0.617}
		&
		& \cellcolor{gray!10}{1.00}
		& \cellcolor{gray!10}{[0.50, 1.99]}
		& \cellcolor{gray!10}{1.000}
		&
		& \cellcolor{gray!10}{0.58}
		& \cellcolor{gray!10}{[0.26, 1.26]}
		& \cellcolor{gray!10}{0.167}\\
		
		No localization
		& 0.80
		& [0.42, 1.54]
		& 0.508
		&
		& 1.00
		& [0.50, 1.99]
		& 1.000
		&
		& 0.50
		& [0.23, 1.08]
		& 0.079\\
		
		\cellcolor{gray!10}{No observed behavior}
		& \cellcolor{gray!10}{1.00}
		& \cellcolor{gray!10}{[0.51, 1.94]}
		& \cellcolor{gray!10}{1.000}
		&
		& \cellcolor{gray!10}{1.00}
		& \cellcolor{gray!10}{[0.50, 1.99]}
		& \cellcolor{gray!10}{1.000}
		&
		& \cellcolor{gray!10}{0.58}
		& \cellcolor{gray!10}{[0.26, 1.26]}
		& \cellcolor{gray!10}{0.167}\\
		
		No suggestions
		& 0.69
		& [0.36, 1.31]
		& 0.255
		&
		& 0.83
		& [0.42, 1.65]
		& 0.603
		&
		& \textbf{0.38}
		& \textbf{[0.18, 0.82]}
		& \textbf{0.013}\\
		
		\cellcolor{gray!10}{No summary}
		& \cellcolor{gray!10}{1.06}
		& \cellcolor{gray!10}{[0.54, 2.07]}
		& \cellcolor{gray!10}{0.865}
		&
		& \cellcolor{gray!10}{1.70}
		& \cellcolor{gray!10}{[0.83, 3.48]}
		& \cellcolor{gray!10}{0.147}
		&
		& \cellcolor{gray!10}{0.85}
		& \cellcolor{gray!10}{[0.38, 1.88]}
		& \cellcolor{gray!10}{0.685}\\
		
		No localization, no suggestions
		& \textbf{0.37}
		& \textbf{[0.20, 0.68]}
		& \textbf{0.002}
		&
		& 0.66
		& [0.34, 1.30]
		& 0.232
		&
		& \textbf{0.22}
		& \textbf{[0.10, 0.46]}
		& \textbf{$<$0.001}\\
		
		\cellcolor{gray!10}{No behavioral context, no localization, no suggestions}
		& \cellcolor{gray!10}{\textbf{0.45}}
		& \cellcolor{gray!10}{\textbf{[0.24, 0.85]}}
		& \cellcolor{gray!10}{\textbf{0.013}}
		&
		& \cellcolor{gray!10}{0.56}
		& \cellcolor{gray!10}{[0.29, 1.10]}
		& \cellcolor{gray!10}{0.092}
		&
		& \cellcolor{gray!10}{\textbf{0.14}}
		& \cellcolor{gray!10}{\textbf{[0.07, 0.30]}}
		& \cellcolor{gray!10}{\textbf{$<$0.001}}\\
		\bottomrule
	\end{tabular}
	\vspace{-0.4cm}
\end{table*}

The last two treatments correspond to more aggressive manipulations, in which multiple information types are removed simultaneously (\eg ``\emph{no localization, no suggestions}''). The reported ORs should be interpreted relative to the ``\emph{no noise}'' treatment, which corresponds to the original issue after removing only text spans that did not convey any of the information types considered in our study. \tabref{tab:ablation_by_model} reports the same analysis separately for each LLM. Since these per-LLM models are fitted on only 65 issues, the available evidence is more limited. We therefore report them for completeness, but center our discussion on the pooled model. Overall, the per-LLM models are consistent with the pooled analysis in the direction of the ORs for the main treatment variables, but not in terms of statistical significance. This is expected given the smaller amount of data available for each LLM.

The first noteworthy finding concerns information types typically considered useful to human developers \cite{Zimmermann:tse2010,Laukannen:esem2011,Soltani:emse2020}, but that appear less critical for AI agents in this setting. Removing behavioral context (\ie reproduction steps and expected behavior), observed behavior, or the issue summary does not significantly affect repair success. The corresponding odds ratios are close to 1.00 in most cases, indicating a limited measurable impact in this controlled setting. This holds both in the pooled model and in the per-LLM models. Still, this does not mean that such information is generally useless. Rather, in information-complete reports, agents can still attempt the repair when these components are removed, as long as additional operational cues, such as localization information and suggested fixes, remain available.

The results differ for localization and suggested fixes. Removing localization alone does not significantly reduce success (OR=0.78, $p$-value=0.203), consistently with the observational analysis. Removing suggested fixes alone, instead, significantly reduces the odds of success (OR=0.65, $p$-value=0.027), confirming that suggested fixes are the most actionable information type for agents.

The greatest degradation occurs when multiple operational cues are removed simultaneously. In particular, removing both localization and suggestions significantly reduces the odds of success, despite all other available information (\ie bug summary, observed behavior, expected behavior, steps to reproduce). In the pooled model, this results in a statistically significant OR of 0.42, and this treatment is also significant (with even lower ORs) for GPT-5-mini and MiniMax M2.5 in the per-LLM models. These results suggest that the most harmful manipulations are those that deprive the agent of both where-to-look and how-to-fix information. 

We also analyze how removing localization cues and suggested fixes affects the cost of agent runs. Specifically, we compare the average cost per fixing attempt in the \emph{no noise} treatment, corresponding to complete issues, with the \emph{no localization, no suggestions} treatment. Although the latter provides shorter issue descriptions and therefore reduces the number of input tokens, the lack of actionable guidance forces agents to take more steps before attempting a fix, increasing overall cost. Indeed, compared to the \emph{no noise} treatment, the average cost per fixing attempt increases by 15\% for GPT-5-mini, accompanied by a 15\% increase in executed steps; by 3\% for Gemini 3 Flash, with a 7\% increase in steps; and by 39\% for MiniMax M2.5, with an 23\% increase in steps.

The control variables confirm that intrinsic task complexity remains a major factor. 
Gold-patch size is negatively associated with success in the pooled model (OR=0.33) and for GPT-5-mini and MiniMax M2.5. This confirms that improving the issue report can make a task more actionable, but it cannot fully compensate for the complexity of the fixes. 

Finally, the \emph{code mention} control is not significant, confirming that the actionable signal is not whether an issue mentions code, but whether code-related information is structured in a way that supports repair, for example, through localization cues or concrete fix suggestions.

Overall, the ablation study refines the observational findings. Human-oriented information, such as summary, observed behavior, and behavioral context, is not independently decisive when operational cues remain available. In contrast, removing suggested fixes significantly reduces success, and removing them together with localization cues causes the largest drop in repair success, showing that agents need at least some guidance on how to approach the repair or where to inspect the code. Thus, agent-ready reports should prioritize concise, repair-directed information over exhaustive descriptions.

\section{Implications \& Lessons Learned}
\label{sub:lessons}

Our results provide several lessons for agentic-first development, where issue reports are not only read by developers but also serve as the initial task specification for repair agents.

\textbf{Information useful for humans is not necessarily decisive for agents.}
Some information types considered important for human-based debugging (\eg steps to reproduce, observed behavior \cite{Zimmermann:tse2010,Laukannen:esem2011,Soltani:emse2020}) appear less decisive for agents. Removing the issue summary, observed behavior, or behavioral context does not significantly affect success when other operational cues remain available. This does not mean that such information is useless. Rather, it suggests that once an issue already contains localization cues or repair guidance, agents rely more on these operational signals than on broader behavioral descriptions.

\textbf{Agentic workflows should be incremental, not one-shot.}
Some of the most useful information types, such as precise localization cues and suggested fixes, may not be available when the issue is first opened. They often emerge later through discussion, debugging, or partial investigation. Some existing agents already support forms of interaction or re-triggering, for example, through explicit labels, mentions, or feedback on generated pull requests. However, these workflows typically require an explicit user action and do not necessarily treat the issue discussion as an evolving task specification that is systematically reprocessed whenever new repair-relevant information is added. Our results suggest that agentic-first workflows should move in this direction. Rather than triggering the agent only once, immediately after issue creation, issue trackers could support incremental re-execution when new comments add relevant information, such as a suspected location, a failing code region, or a possible fix. This would turn issue tracking into an iterative human-agent process, where developers and users progressively enrich the issue until it becomes actionable for the agent.

\textbf{Localization may need to be automatically derived.}
Although not significant on their own, localization cues become critical when suggested fixes are missing, as shown by our ablation study. This suggests that agentic-first development should not rely solely on reporters to identify the bug's location.
In many cases, users opening an issue may not know the relevant file, function, or line. Therefore, agentic repair systems may benefit from a multi-agent or multi-stage design in which a specialized localization component first identifies suspicious code regions, and a repair agent then uses this information to generate a patch. Such localization could be produced by static analysis \cite{Ayewah:ieee2008, Bessey:cacm2010}, information retrieval \cite{Zhou:icse2012, Saha:ase2013, Ye:fse2014}, test execution \cite{Jones:ase2005,Abreu:jss2009}, stack-trace analysis \cite{Wu:issta2014, Moreno:icsme2014}, or a dedicated LLM-based localization agent \cite{xia2025agentless,Xu:tse2025,Qin:tse2025}. Our results suggest that improving the agent's starting point in the codebase can be as important as improving the patch-generation step itself.

\textbf{Issue templates should be treated as agent-specific interfaces.}
The per-LLM analyses show that different backbones do not exploit issue information in the same way. Natural-language suggested fixes are robustly useful across models in the observational analysis, but other information types show model-specific patterns. This means that there may not be a single universal template for agent-ready issues. Project owners should treat issue templates as interfaces between humans and the specific agent used in the project. 

\textbf{References should be summarized and properly contextualized, not merely linked.}
The negative association observed for references in the observational analysis should not be interpreted as evidence that references are inherently harmful. A more plausible explanation is that references often indicate issues whose relevant context is distributed across external discussions, previous reports, documentation, or commits. Humans figure out what references are for, decide whether to inspect them, and determine which parts are relevant. Agents, however, may not reliably identify the repair-relevant information unless it is explicitly surfaced in the issue text. Thus, in agentic-first reporting, references should be accompanied by short summaries explaining why they are relevant and what information the agent can extract from them.

% !TEX root = ../main.tex
%%%%%%%%%%%%%%%%%%%%%%%%%%%%%%%%%%%%%%%%
%%%%%%%%%%%%%%%%%%%%%%%%%%%%%%%%%%%%%%%%
\section{Related Work} \label{sec:related}
%%%%%%%%%%%%%%%%%%%%%%%%%%%%%%%%%%%%%%%%
%%%%%%%%%%%%%%%%%%%%%%%%%%%%%%%%%%%%%%%%
We discuss the literature related to (i) works looking at bug report quality and information content, and (ii) automated program repair at the repository level.

\subsection{Bug report quality and information content}
What makes a bug report useful has been studied primarily from developers' perspectives. Developer surveys consistently identify reproduction steps, stack traces, observed and expected behavior, and test cases as the most valuable report elements, while severity and environment details are typically valued less \cite{Zimmermann:tse2010,Laukannen:esem2011,Soltani:emse2020}. Laukkanen and M\"antyl\"a \cite{Laukannen:esem2011} reproduced Zimmermann \etal's bug-report survey \cite{Zimmermann:tse2010} in an industrial context by surveying 74 developers from six software organizations. 

They found that developers most often considered steps to reproduce, screenshots, and the affected part of the application useful for fixing defects, with steps to reproduce being by far the most important item. This finding was consistent with the original study.

What developers value, however, is not always what reporters find easy to supply \cite{Zimmermann:tse2010,Sasso:qrs2016}. In practice, bug reports are often incomplete, ambiguous, or incorrect \cite{Davies:esem2014}, partly because reporters and developers do not share the same knowledge of the system \cite{Huo:icsme2014}. Indeed, errors in reproduction steps and missing information are considered among the most severe problems for developers \cite{Zimmermann:tse2010} and are also leading causes of non-reproducible reports \cite{erfani:msr2014}.

Issue templates address this problem by eliciting relevant elements upfront and standardizing report descriptions, although their adoption may also reduce submission volume and increase the reporting burden when templates request too much information \cite{sulun:tosem2024,Li:tse2023,Li:jss2023}.
These findings have motivated the development of techniques to detect missing or low-quality information and to help reporters improve it. Existing approaches range from feature-based classifiers \cite{Zimmermann:tse2010}
to element-level detectors for specific information types such as reproduction steps and expected behavior \cite{Chaparro:fse2017,Chaparro:fse2019,Song:fse2020},
to tools that assist reporters at submission time \cite{Moran:fse2015, Song:icse2023} 
and, more recently, to LLM-based techniques that assess and regenerate missing content \cite{Bo:ase2024,Feng:icse2024,Mahmud:icpc2025,Akyol:ease2026}.

Overall, prior work indicates what makes a bug report useful for humans. Our work asks a complementary question: which of these information types makes a report actionable for software repair agents?

\subsection{Automated Program Repair \& Repository-Level Agents}
Over the years, Automated Program Repair (APR) has evolved from approaches based on manually designed or automatically mined fix templates to learning-based techniques relying on a variety of machine learning models \cite{Zhang:tosem2023survey}. More recently, the field has shifted increasingly toward LLM-based repair, ranging from one-shot patch generation to iterative refinement and, most recently, agentic formulations. A systematic literature review by Zhang \etal \cite{Zhang:tosem2026systematic} confirms both the growing prevalence of LLM-based approaches and the emergence of agent-based ones, where an agent autonomously navigates a repository, localizes the fault, generates a patch, and runs the test suite with minimal human intervention.

These agents rely on architectures ranging from rigid pipelines to fully autonomous loops \cite{rombaut2026}.
At one end of this spectrum, fixed procedural pipelines decompose repair into predefined stages, such as fault localization, patch generation, and validation, without delegating control over the process to the model \cite{xia2025agentless,Zhang:2024,Ruan:icse2025}. 
At the other end, ReAct-style agents \cite{yao2023react} give the model broad autonomy within a perceive-act loop, allowing it to interact with shell commands and editing tools to inspect the repository, modify code, and validate candidate fixes \cite{yang:neurips2024,wang:iclr2025}. This latter paradigm is also reflected in industrial CLI agents \cite{codexcli,geminicli,opencode}.

A growing body of work augments these scaffolds with test-time search and ensembling \cite{antoniades2025swesearch,aggarwal2025dars,traeresearchteam2025trae}, multi-agent decomposition \cite{Tao:nips2024,chen2024coder,wadhwa2024masai}, and experience reuse or self-evolution \cite{xia2025live,chen2026sweexp,mu2026experepair}.

All such agents are conditioned on the issue report they receive, and only a few studies have examined how its content relates to resolution. Meng \etal \cite{meng2026llm} assess the quality of information content in \swebench Verified issues by using LLM-assigned scores and look for a relationships between such scores and fixing success by six repair systems. They found that the quality of solution and localization information has the highest impact on the likelihood of fixing success. More recently, Ehsani \etal \cite{ehsani2026llms} showed that LLMs repairing a given buggy function succeed when they attend to bug descriptions, stack traces, and tests, and fail when they over-attend to version information. These studies rely on automatic scores or on model attention. We instead manually annotate report content at the character level, estimate the association of each information type with agent success while controlling for confounding factors, and confirm it with an ablation study on the same reports.

% !TEX root = ../main.tex

%%%%%%%%%%%%%%%%%%%%%%%%%%%%%%%%%%%%%%%%
%%%%%%%%%%%%%%%%%%%%%%%%%%%%%%%%%%%%%%%%
\section{Threats to Validity} \label{sec:threats}
%%%%%%%%%%%%%%%%%%%%%%%%%%%%%%%%%%%%%%%%

%%%%%%%%%%%%%%%%%%%%%%%%%%%%%%%%%%%%%%%%

Threats to \textbf{construct validity} concern the relationship between theory and observation. One admitted limitation of the study is that we consider the presence of certain types of information in the issue report as binary variables, while it is known that, for example, not all expected behavior, observed behavior, steps to reproduce, or any other piece of information may be written with the same level of detail and quality. Similarly, a suggested fix may describe the complete solution or only part of it. To better understand the impact of the completeness of such information, two authors compared each of the 114 suggested fixes with the gold patch (Cohen $k$=0.92 \cite{kappa}) and found that 68 describe the complete solution, 29 a partial one, 13 an alternative solution, and 4 are misleading \cite{replication}. In the ablation study, removing a complete suggested fix strongly decreases the chances of a successful fixing (-11\%), whereas removing a partial one leaves it unchanged. The benefit of suggested fixes therefore comes mainly from reports that already describe a complete solution at issue-opening time.

A further threat may concern the subjectivity or imprecision in the bug report annotation quality. We mitigated this threat by employing multiple, independent annotators. Lastly, the ablation study artificially removes pieces of information, yet this may not correspond to how humans would have reported the same issue (\eg more naturally) without that information.

Threats to \textbf{internal validity} concern factors internal to our study that can influence our results. Our study is observational and seeks correlations, not causation. We mitigated the presence of confounding effects through (i) the issue random effect, which also captures repository-level differences, and (ii) control variables related to the issue and patch size.
However, there could be further confounding factors we did not consider. As explained in \secref{sec:design}, we used redundant variable analysis to avoid collinearity among variables. 

At the same time, in the ablation study, we did not consider all possible information combinations or their interactions, as such a number would not be tractable, given the dataset size.

Agents could also retrieve external information referenced by the issues. They ran in the standard \swebench environment, where network access is enabled to install dependencies but the repository contains only the snapshot preceding the fix, without its history. We inspected all 8,649 trajectories for commands that retrieve remote content (\eg curl, wget, git clone, fetch, or pull). Such commands appear in 55 runs (0.6\%), mostly to reproduce the bug, and only one run accessed a resource referenced by the issue (django-11555), which the same model solved in all its other runs without accessing it.

Threats to \textbf{conclusion validity} concern the extent to which our conclusions can be statistically supported. Given the relatively small data set and the number of considered variables, we may have incurred a type-II error for some variable where we did not observe statistically significant differences.

Threats to \textbf{external validity} concern the generalizability of our findings. \swebenchverified includes bug reports from 12 repositories (11 in our ablation subset). However, bug reports from certain ecosystems may follow a template we did not consider, for which we did not fully explore the LLM's ability to leverage them. Furthermore, we consider one agentic AI with three different LLMs; our conclusion may not generalize to other agentic AI and models. 
% !TEX root = main.tex
%%%%%%%%%%%%%%%%%%%%%%%%%%%%%%%%%%%%%%%%
%%%%%%%%%%%%%%%%%%%%%%%%%%%%%%%%%%%%%%%%
\section{Conclusion and Future Work} \label{sec:conclusion}

We studied what makes a bug report \emph{agent-ready}. Starting from \swebenchverified, we manually classified issues by change type and annotated bug report sentences based on the information they convey. We then evaluated the ability of mini-swe-agent, powered by three LLM backbones, to resolve the annotated bug reports and modeled the relationship between report information and agent success.

Our results show that agents benefit most from information that enables the repair task. In particular, suggested fixes are positively associated with successful repairs, while localization cues become critical when suggested fixes are missing.
Conversely, information traditionally emphasized in human-oriented bug reporting, such as steps to reproduce, appears less decisive for agents when more actionable cues remain available. Our sentence-level ablation study further supports this finding. Removing most individual report components has a limited impact, whereas removing suggested fixes reduces agent effectiveness, especially when localization cues are also removed.

These findings suggest that issue reports should be reconsidered as interfaces between humans and repair agents. Future research should explore multi-stage agent architectures in which dedicated localization or issue-understanding agents enrich reports before patch generation.

\balance
\bibliography{main}
\bibliographystyle{IEEEtran}

\end{document}